\documentclass[referee]{raa}            

\newcommand{\kep}{{\it Kepler }}
\newcommand{\corot}{{\it CoRoT }}
\newcommand{\tess}{{\it TESS }}
\newcommand{\gaia}{{\it Gaia }}

\usepackage{graphicx,times}             
\usepackage{natbib}
\usepackage{amssymb,amsmath}
\usepackage{threeparttable}
\usepackage{booktabs}

\bibpunct{(}{)}{;}{a}{}{,}

\begin{document}

   \title{KIC 10417986: Spectroscopic confirmation of the nature of the binary system with a \bm{\delta} Scuti component}
   \volnopage{Vol.0 (20xx) No.0, 000--000}     
   \setcounter{page}{1}          

   \author{Guo-Jie Feng
      \inst{1,2}
   \and Ali Esamdin 
      \inst{1,2} $^{\star}$
   \and Jian-Ning Fu 
      \inst{3} $^{\dagger}$
   \and Hu-Biao Niu
      \inst{1}
   \and Peng Zong
      \inst{3}
   \and Tao-Zhi Yang
      \inst{4}
   \and Shu-Guo Ma
      \inst{1,2}
   \and Jing Xu
      \inst{1}
   \and Chun-Hai Bai
      \inst{1}
   \and Yong Wang
      \inst{1}
   \and Wei-Chao Sun
      \inst{1}
   \and Xin-Liang Wang
      \inst{1}
   }
   \institute{Xinjiang Astronomical Observatory, Chinese Academy of Sciences, Urumqi, Xinjiang 830011, P.R.China; $^{\star}${\it aliyi@xao.ac.cn}\\
	\and University of Chinese Academy of Sciences, No.19(A) Yuquan Road, Shijingshan District, Beijing 100049, P.R.China; \\
	\and Department of Astronomy, Beijing Normal University, Beijing 100875, P.R.China; $^{\dagger}${\it jnfu@bnu.edu.cn}\\
	\and Ministry of Education Key Laboratory for Nonequilibrium Synthesis and Modulation of Condensed Matter, School of Physics, Xi’an Jiaotong University, 710049 Xi’an, P.R.China; \\
\vs\no
   {\small Received~~20xx month day; accepted~~20xx~~month day}}

\abstract{ KIC 10417986 is a short orbital period (0.0737 d) ellipsoidal variable star with a $\delta$ Sct and $\gamma$ Dor hybrid pulsations component discovered by \kep. The ground-based spectroscopic observations were carried out in the winters of 2020 and 2021 to investigate the binary nature of this star. We derive the orbital parameters using the rvfit code with a result of \emph{K}$_1$ = 29.7 $\pm$ 1.5 kms$^{-1}$, $\gamma$ = $-$18.7 $\pm$ 1.7 kms$^{-1}$, and confirm an orbital period of 0.84495 d instead of the result given by \kep. The atmospheric parameters of the primary are determined by the synthetic spectra fitting technique with the estimated values of \emph{T}$_{eff}$ = 7411 $\pm$ 187 K, log \emph{g} = 4.2 $\pm$ 0.3 dex, [M/H] = 0.08 $\pm$ 0.09 dex and \emph{v}sin\emph{i} = 52 $\pm$ 11 km/s. KIC 10417986 is a circular orbit binary system. From the single-lined nature and mass function of the star, the derived orbital inclination is 26 $\pm$ 6$^{\circ}$, and the mass of the secondary is $0.52^{+0.18}_{-0.09}$ \emph{M}$_\odot$, which should be a late-K to early-M type star. Fourteen frequencies are extracted from \kep light curves, of which six independent frequencies in the high-frequency region are identified as the $p$-mode pulsations of $\delta$ Sct star, and one independent frequency in the low-frequency region ($f_{2}$ = 1.3033 cd$^{-1}$) is probably the rotational frequency due to the starspots rather than the ellipsoidal effect.
}

   \authorrunning{G.-J. Feng et al.}            
   \titlerunning{KIC 10417986: SB1 with a $\delta$ Scuti component}  

   \maketitle

%
%
\section{Introduction}\label{sec:Int}
The study of binary stars, especially eclipsing binaries has gained a new prospect since the beginning of the \corot, \kep, and \tess space missions \citep{2010ASPC..435...41D, 2011A&A...534A.125U, 2013A&A...552A..60M, 2016A&A...594A.100B, 2021AAS...23753001M}. The high-precision photometric data from space, combined with ground-based spectroscopic data, allows a precise determination of stellar properties, such as mass, radius, and luminosity \citep[e.g.,][]{2016ApJ...826...69G, 2019ApJ...887..253C, 2020ApJ...895..136C, 2020A&A...642A..91L}. In particular, by studying pulsating components in eclipsing binary systems, such as the $\delta$ Sct and $\gamma$ Dor variables , one can build appropriate asteroseismic models based on the precise stellar parameters to constrain the stellar structure and evolution. However, in the case of non-eclipsing binary systems, the two stars do not eclipse each other in the direction of our sight line. Especially for single-line spectroscopic binaries, whose mass ratio cannot be determined \citep{2022AJ....163..220V}, this lack of information will limit the method of detailed asteroseismic modeling of the pulsating components.

$\delta$ Sct are stars located on what is known as the instability strip, on or a little above the main sequence in the Herzsprung-Russell (H-R) diagram \citep{2000ASPC..210....3B}. They are late-A to early-F spectral type stars with masses between 1.5 and 2.5 $M_{\odot}$, and luminosities in the range 0.6 $\leq$ log($L/L_{\odot}$) $\leq$ 2.0. These stars typically pulsate in radial and low-order non-radial modes, belong to $p$-mode pulsations with amplitudes less than 0.3 mag, and pulsating periods range from 15 min up to 8 h \citep{2001A&A...366..178R, 2011A&A...534A.125U}. The oscillations of those stars are self-excited through the $\kappa$ mechanism that occurs in the partial ionization zone of He II \citep{1994A&A...286..815L, 1995ARA&A..33...75G, 2000ASPC..210....3B, 2009AIPC.1170..403H, 2015MNRAS.452.3073B}. They also pulsate in high-order non-radial modes, which may be due to the turbulent pressure in the Hydrogen convective zone\citep{2014ApJ...796..118A}.

KIC 10417986 = HD 187276 ($K_{p} = 9.13$ mag, $\alpha_{J2000}=19^{h}46^{m}59.70^{s}$, $\delta_{J2000}=+47^{\circ}32^{'}37.71^{''}$) was identified as a binary star with a very short orbital period of 0.0737(1) days and suggested to be a possible triple system \citep{2014AJ....147...45C, 2016AJ....151...68K}. Then \cite{2019A&A...630A.106G} flagged it as an ellipsoidal variable star with $\delta$ Sct and $\gamma$ Dor hybrid pulsations by applying a data inspection tool (DIT) to the whole \kep light curves. Based on known parameters from \kep and \gaia catalogs (see Table \ref{tab:stellar parameters}), the primary component is a late-A or early-F spectral type star.

KIC 10417986, as an example for astronomical research at Chiang Rai Rajabhat University, was studied for the first time by \cite{2021JPhCS1719a2017A}, who derived the absolute parameters of two components under the assumption that the system is a contact binary system whose components are main-sequence stars. They found that the system might have a high mass transfer rate from the primary component onto the secondary one. However, this study was based on the light curves of \kep and \tess online archives without spectral data, and no detailed study of the system was possible. In this paper, we carry out a more detailed study of KIC 10417986 based on the \kep photometric data and the ground-based spectroscopic data, which are described in Section \ref{sec:P&S}. The orbital parameters, atmospheric parameters, and frequency analysis are presented in Section \ref{sec:analysis}. We show a brief discussion on the probable rotation frequency $f_{2}$, orbital inclination, the mass of the secondary, and the age of the system in Section \ref{section:DISCUSSION}. The summary of this work is given in Section \ref{section:SUMMARY}.

\begin{table*}
\begin{center}	
\caption{Table of stellar parameters from catalogs for KIC 10417986.}
\label{tab:stellar parameters}
\begin{tabular}{cccc}
\hline
\hline
Parameters & The Spitzer \kep Survey Catalog & \tess Input Catalog & \gaia DR2  \\
		  & {\citep{2021ApJS..254...11W}} & {\citep{2019AJ....158..138S}} & {\citep{2018A&A...616A...1G}}\\
\hline
\rule{0pt}{9pt}
ID                       & KIC 10417986              & TIC 272599738          & 2086579087597378816 \\
$T_{eff}$ (K)            & 7430 $\pm$ 324            & 7808 $\pm$ 143         & 7484 \\
log g (dex)              & 4.218 $\pm$ 0.210         & 4.1539 $\pm$ 0.0751    &  \\
$[$Fe/H$]$ (dex)         & 0.07 $\pm$ 0.35           &                        &  \\
Mass ($M_{\odot}$)       & 1.581 $\pm$  0.214        & 1.853 $\pm$ 0.291      &  \\
Radius ($R_{\odot}$)     & $1.619^{+0.575}_{-0.230}$ & 1.888 $\pm$ 0.053      & 1.95 \\
Parallax (mas)           &                           &                        & 4.1128 $\pm$ 0.0302  \\
Distance (pc)            &                           & 241.4630 $\pm$ 1.7760  & $241.4635^{+1.7886}_{-1.7631}$ \\
Luminosity ($L_{\odot}$) & 				            & 11.93847 $\pm$ 0.47378 & $10.774^{+0.127}_{-0.126}$     \\
\hline
\end{tabular}
\end{center}
\end{table*}

\section{Kepler Photometric Data and Ground-based Spectroscopic Data}\label{sec:P&S}

\subsection{\bm{\kep} photometric data}

\begin{figure*}
\centering
\includegraphics[width=1\textwidth]{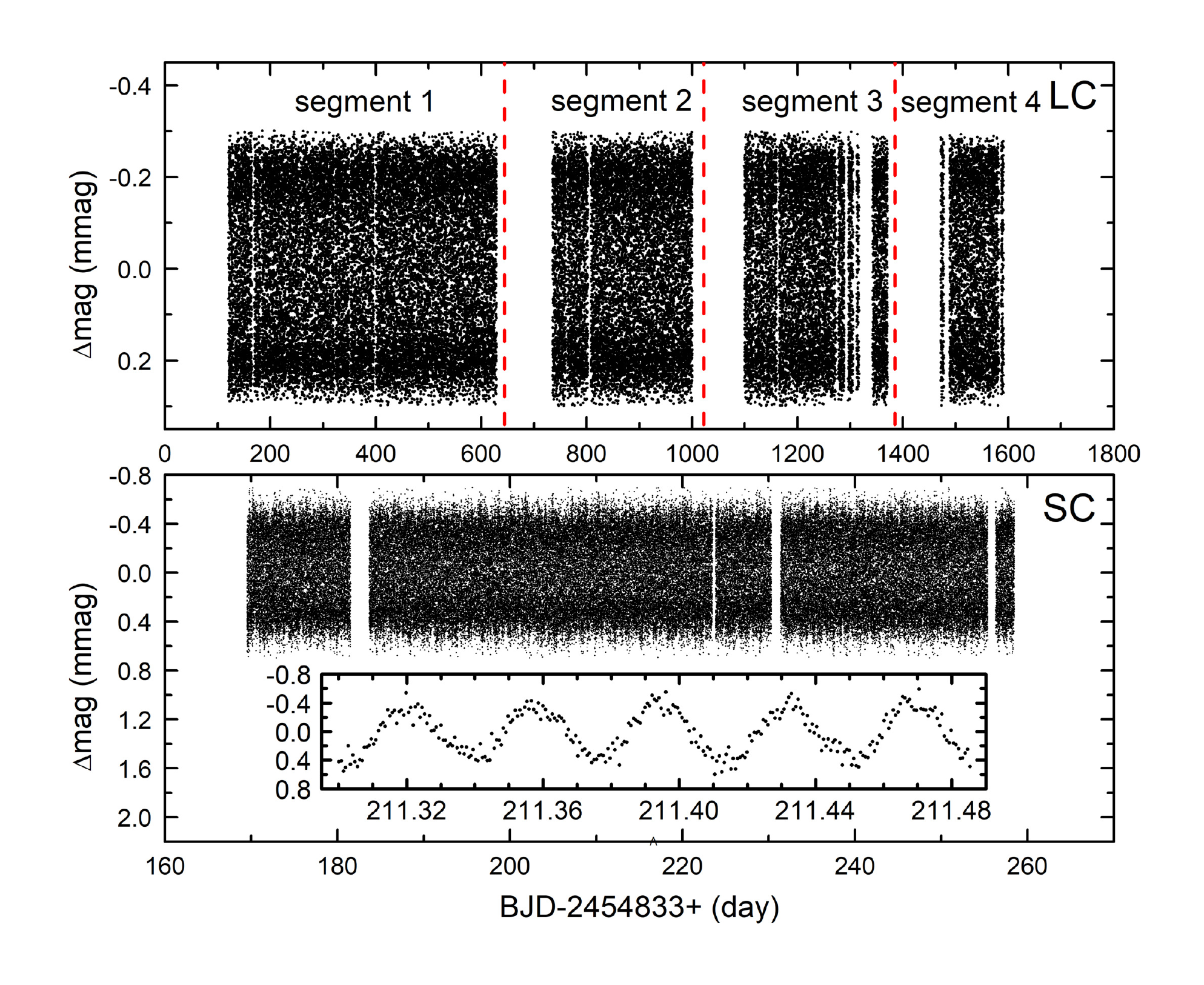}
\caption{Detrended \kep LC (top panel) and SC (bottom panel) light curves of KIC 10417986. The inset of SC shows the zoomed area covering 0.185 days of obervations to visualize the photometric variability.}
\label{fig:LC-SC-ligthcurves}
\end{figure*}

KIC 10417986 was observed by NASA space mission \kep from BJD 2454953.54 to 2456424.00, including fifteen quarters (Q0-Q6, Q8-Q10, Q12-Q14, and Q16-Q17) of long cadence (LC) data with 29.4 minutes sampling and one quarter (Q2.1, Q2.2, and Q2.3) of short cadence (SC) data with 58.8 seconds sampling \citep{2011AJ....141...83P, 2011AJ....142..160S, 2016AJ....151...68K}. This star's target pixel files (TPFs) were downloaded from Mikulski Archive for Space Telescopes (MAST\footnote{https://archive.stsci.edu/}), and the Lightkurve package \citep{2018zndo...1181929V, 2019AAS...23344501D, 2021zndo...1181928B} was adopted to extract light curves from the TPFs. A series of different apertures of pixel sizes were tested on TPFs to optimize the photometry. After extracting the light curves, we detrended and normalized the flux by a linear or polynomial fitting. Then, we converted the corrected flux to magnitude. Finally, a total of 50168 LC data points and 123255 SC data points were obtained after removing outliers by sigma clipping. Both LC and SC light curves are shown in Figure \ref{fig:LC-SC-ligthcurves}. It can be seen that the period of the maximum amplitude is close to 0.04 days, and the mean amplitude ($\sim$0.6 mmag) of the SC light curve is larger than that ($\sim$0.3 mmag) of the LC light curve because the high-frequency amplitude of LC data is suppressed by the long exposure time \citep{2018ApJ...863..195Y}.

\subsection{Ground-based Spectroscopic Data}

\begin{table*}
\centering
\caption{Journal of spectroscopic observations for KIC 10417986.}
\label{tab:spec obs}
\begin{tabular}{ccccccc}
\hline
\hline
Date & Observatory & Telescope & Instrument & Spectra & Exposure time & S/N\\
\hline
2020 Oct.4  & NAOC & 2.16-m & BFOSC (E9+G11+slit1.6$''$) & 8 & 15 min  &  340\\
2020 Oct.5  & NAOC & 2.16-m & BFOSC (E9+G11+slit1.6$''$) & 8 & 15 min  &  150\\
2021 Oct.20 & NAOC & 2.16-m & BFOSC (E9+G11+slit1.6$''$) & 2 & 30 min  &  710\\
2021 Nov.24 & XAO  & 1.2-m  & NES (low-resolution)       & 3 & 30 min  &  39 \\
2021 Nov.25 & XAO  & 1.2-m  & NES (moderate-resolution)  & 3 & 60 min  &  17 \\
2021 Nov.29 & XAO  & 1.2-m  & NES (moderate-resolution)  & 3 & 60 min  &  48 \\
2022 Mar.26 & NAOC & 2.16-m & HRS                        & 2 & 30 min  &  26 \\
\hline
\end{tabular}
\end{table*}

The spectroscopic observations for KIC 10417986 aimed to derive the radial velocities (RVs) of the components. As a first step, we have gathered 16 echelle spectra with Beijing Faint Object Spectrograph and Camera (BFOSC) attached to the 2.16-m telescope at the Xinglong Station of National Astronomical Observatories, Chinese Academy of Sciences (NAOC) on October 4 and 5 of 2020. We chosed the combination of echelle and grism E9+G11 with the slit size of 1.6$''$, which provided a resolution of $\sim$0.55 {\rm \AA}pixel$^{-1}$ and a wavelength coverage from 3900 to 9000 \AA \citep{2016PASP..128k5005F}. Eight spectra per night were taken with the goal to cover the possible orbital period of 0.0737 days ($\approx$ 2 hours) suggested by \cite{2016AJ....151...68K}. The exposure time for each spectrum was 15 min with a temporal resolution $\sim$12.5\% of the orbital period. However, the result showed that the RVs do not change periodically for 2 hours in the two-hour observation every night, and the RVs difference between the two days is about 30 km/s.

For solving this system, supplemented spectroscopic observations were performed with the BFOSC instrument of the 2.16-m telescope at the Xinglong Station of NAOC, and the Nanshan Echelle Spectrograph (NES) instrument of the 1.2-m telescope at the Nanshan Station of Xinjiang Astronomical Observatory (XAO) in the winter of 2021. The Nanshan Echelle Spectrograph is a fiber-fed moderate- and low-resolution echelle spectrograph with 2.4$''$ aperture fiber, which has been developed by Astronomical Consultants \& Equipment, Inc. (ACE). The NES instrument using the low-resolution grating (95 grooves mm$^{-1}$) and moderate-resolution grating (79 grooves mm$^{-1}$) was employed for the observations. The low-resolution grating provided a spectral coverage from 3680 to 10290 {\rm \AA} while the resolution of the spectra is from 0.13 to 0.35 {\rm \AA}pixel$^{-1}$, and the moderate-resolution grating provided a spectral coverage from 3770 to 10040 {\rm \AA} while the resolution of the spectra is from 0.069 to 0.178 {\rm \AA}pixel$^{-1}$. The observations were carried out using the BFOSC (E9+G11+slit1.6$''$) on October 20, the NES (low-resolution) on November 24, the NES (moderate-resolution) on 25, and the NES (moderate-resolution) with 2x2 on-chip binning on 29. A total of eleven spectra were obtained during this observation run.

High-resolution spectroscopic observations for KIC 10417986 were also carried out with the fiber-fed High-Resolution Spectrograph (HRS) \citep{2016PASP..128k5005F} instrument attached to the 2.16-m telescope at the Xinglong Station, National Astronomical Observatories, Chinese Academy of Sciences (NAOC) on March 26 of 2022. The spectrograph is equipped with a back-illuminated red-sensitive 4096 $\times$ 4096 E2V CCD-203-82 with a pixel size of 12 $\mu$m. This set-up provides a resolution of $\sim$0.025 {\rm \AA}pixel$^{-1}$ corresponding to 2.4$''$ aperture fiber and a spectral coverage range from 3600 to 10000 \AA. We have collected 2 high-resolution spectra with an exposure time of 30 min. The signal-to-noise ratio is about 26 for each spectrum.

A detailed journal of all spectroscopic observations for KIC 10417986 is given in Table \ref{tab:spec obs}, and a total of 29 spectra are obtained. Firstly, the spectra were reduced by the IRAF/CCDRED package with bias subtraction, trimming, and flat-field correction. Then, we used the STSDAS/lacos\_spec package \citep{2001PASP..113.1420V} to remove the cosmic rays. Finally, the extraction and normalization of spectra were done using the IRAF/ECHELLE package. Meanwhile, wavelength calibration was performed using the Fe-Ar or Th-Ar lamp taken during the same observation night.

Considering the limit of resolution and S/N of our spectra, we did not find any clue of spectral lines belonging to the secondary component of KIC 10417986. This indicates that the luminosity of the secondary is much lower than that of the primary.

\section{Analysis}\label{sec:analysis}
\subsection{Determination of the spectroscopic orbital solution}

\begin{table}
\centering
\caption{Measured RVs for the primary of KIC 10417986.}
\label{tab:RVs}
\setlength{\tabcolsep}{10mm}{
\begin{tabular}{cccc}
\hline
\hline
HJD & Orbital & RV & RV Error \\
(2459000+) & Phase & (kms$^{-1}$) & (kms$^{-1}$) \\
\hline
126.98428 & 0.77590 & 6.2   & 5.9      \\
126.99476 & 0.78830 & 8.5   & 7.7      \\
127.00523 & 0.80070 & 4.8   & 6.4      \\
127.01570 & 0.81309 & 5.6   & 6.6      \\
127.02618 & 0.82549 & 0.3   & 6.9      \\
127.03665 & 0.83788 & $-$1.0  & 6.1      \\
127.04712 & 0.85028 & $-$2.9  & 5.0      \\
127.05760 & 0.86267 & 0.6   & 5.8      \\
128.03107 & 0.01478 & $-$24.3 & 4.1      \\
128.04154 & 0.02718 & $-$19.0 & 6.0      \\
128.05202 & 0.03957 & $-$27.5 & 5.2      \\
128.06249 & 0.05197 & $-$26.9 & 7.2      \\
128.07296 & 0.06436 & $-$29.6 & 1.6      \\
128.08344 & 0.07676 & $-$32.7 & 5.0      \\
128.09391 & 0.08916 & $-$28.5 & 6.4      \\
128.10439 & 0.10155 & $-$30.3 & 5.6      \\
508.04467 & 0.76163 & 11.9  & 2.5      \\
508.06559 & 0.78638 & 13.3  & 2.0      \\
543.03243 & 0.16971 & $-$47.6 & 7.6      \\
543.05401 & 0.19526 & $-$51.8 & 3.7      \\
543.09270 & 0.24105 & $-$54.9 & 12.0     \\
544.02667 & 0.34640 & $-$47.5 & 6.9      \\
544.08042 & 0.41001 & $-$34.5 & 11.3     \\
544.12251 & 0.45983 & $-$21.3 & 8.4      \\
548.04498 & 0.10207 & $-$41.4 & 10.5     \\
548.08786 & 0.15282 & $-$45.0 & 8.1      \\
548.12972 & 0.20237 & $-$45.5 & 3.2      \\
665.33057 & 0.90980 & $-$3.0  & 2.9      \\
665.35296 & 0.93631 & $-$6.2  & 5.0      \\
\hline
\end{tabular}
}
\end{table}

\begin{table}
\centering
\caption{Orbital parameters of KIC 10417986.}
\label{tab:fitRVs}
\setlength{\tabcolsep}{15mm}{
\begin{threeparttable}
\begin{tabular}{lr}
\hline
\hline
Parameter                       &Value\\
\hline
\multicolumn{2}{c}{Adjusted Quantities}\\
\hline
$P$ (d)          &0.84495 $\pm$ 0.00002       \\
$T_p$ (HJD)      &2459126.3287 $\pm$ 0.0095 \\
$e^a$            &0.0                       \\
$\omega^a$ (deg) &90.0                      \\
$\gamma$ (km/s)  &$-$18.7 $\pm$ 1.7           \\
$K_1$ (km/s)     & 29.7 $\pm$ 1.5           \\
\hline
\multicolumn{2}{c}{Derived Quantities}\\
\hline
$a_1\sin i$ ($R_\odot$)  &0.50 $\pm$ 0.02       \\
$f(m_1,m_2)$ ($M_\odot$) &0.0023 $\pm$ 0.0003 \\
\hline
\multicolumn{2}{c}{Other Quantities}\\
\hline
$\chi^2$                &16.799296  \\
$N_{obs}$ (primary)     &29         \\
Time span (days)        &538.36868  \\
$rms_1$ (km/s)          &4.2       \\
\hline
\end{tabular}

     \begin{tablenotes}
        \footnotesize
        \item $^a$ Parameter fixed beforehand.
     \end{tablenotes}
\end{threeparttable}
}
\end{table}

In order to extract the RVs of KIC 10417986 from the reduced spectra, the following formula is used:
\begin{equation}
\label{eq:RV}
RV = C\times \frac{\Delta \lambda }{\lambda }
\end{equation}
Where \emph{C} is the speed of light in a vacuum, $\lambda$ is the standard center wavelength of the spectral absorption line, and $\Delta$$\lambda$ is the center wavelength difference between the object and the RV standard stars for the same absorption line. For KIC 10417986, we can only measure the RVs of the primary limited by the spectrograph resolution and luminosity difference between the primary and secondary stars. Firstly, we measure the center positions of the Balmer lines (\emph{H}$_\alpha$, \emph{H}$_\beta$, \emph{H}$_\gamma$, and \emph{H}$_\delta$) by using a Gaussian fit. Then, the calibration of the RVs is performed with the RV standard star HR 328 or HIP 087998 observed during the same night. Finally, the heliocentric velocity is calculated by using the IRAF/RVSAO/\emph{bcvcorr} package to correct the motion of Earth. The mean values of the derived RVs from our spectra are listed in Table \ref{tab:RVs}.

We subsequently determine the orbital parameters of KIC 10417986 by using the rvfit\footnote{http://www.cefca.es/people/~riglesias/rvfit.html} code \citep{2015PASP..127..567I}, which uses Adaptive Simulated Annealing (ASA) global minimization method to fit radial velocities of binaries. In the beginning, we set the orbit period $P$, the time of periastron passage $T_p$, the orbital eccentricity $e$, the argument of periastron $\omega$, the systemic radial velocity $\gamma$, and the the radial velocity amplitude $K_1$ of the primary as free parameters in the first run. The results show a very small value of eccentricity $e$, indicating a circular orbit. So we fix the eccentricity $e$ to zero and $\omega$ to 90.0 degrees in the second run. The same results are fitted to the rest of the parameters. The orbital phases are listed in the second column of Table \ref{tab:RVs}. The adjusted and derived orbital parameters are listed in Table \ref{tab:fitRVs}, and the uncertainties are obtained with Markov Chain Monte Carlo (MCMC) analysis.

\begin{figure}
\centering
\includegraphics[width=1\textwidth]{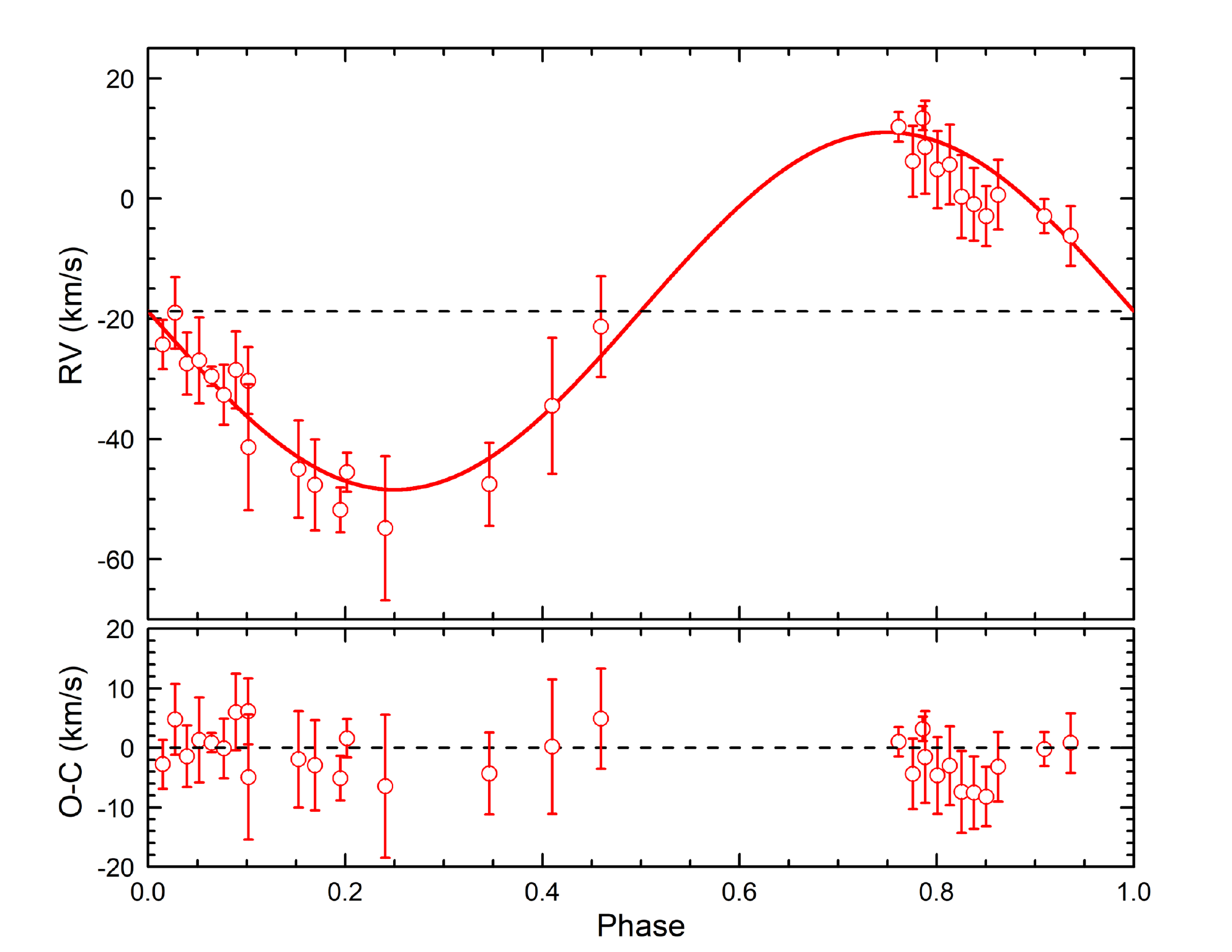}
\caption{Top panel: The phase diagram of radial velocities for the primary of KIC 10417986 with the observed RVs (red open circles) and the orbital fitting (red solid lines). Bottom panel: residuals with uncertainties.}
\label{fig:RV}
\end{figure}

The fitted radial velocity curve compared with the RV measurements of the primary is shown in Figure \ref{fig:RV}. The red solid line in the top panel shows the best-fitting with an orbital period of $P$ = 0.84495 d, velocity variation amplitude of \emph{K}$_1$ = 29.7 kms$^{-1}$, and the systemic radial velocity of $\gamma$ = $-$18.7 kms$^{-1}$. The bottom panel shows the residuals of the best-fitting. 

\begin{figure}
\centering
\includegraphics[width=1\textwidth]{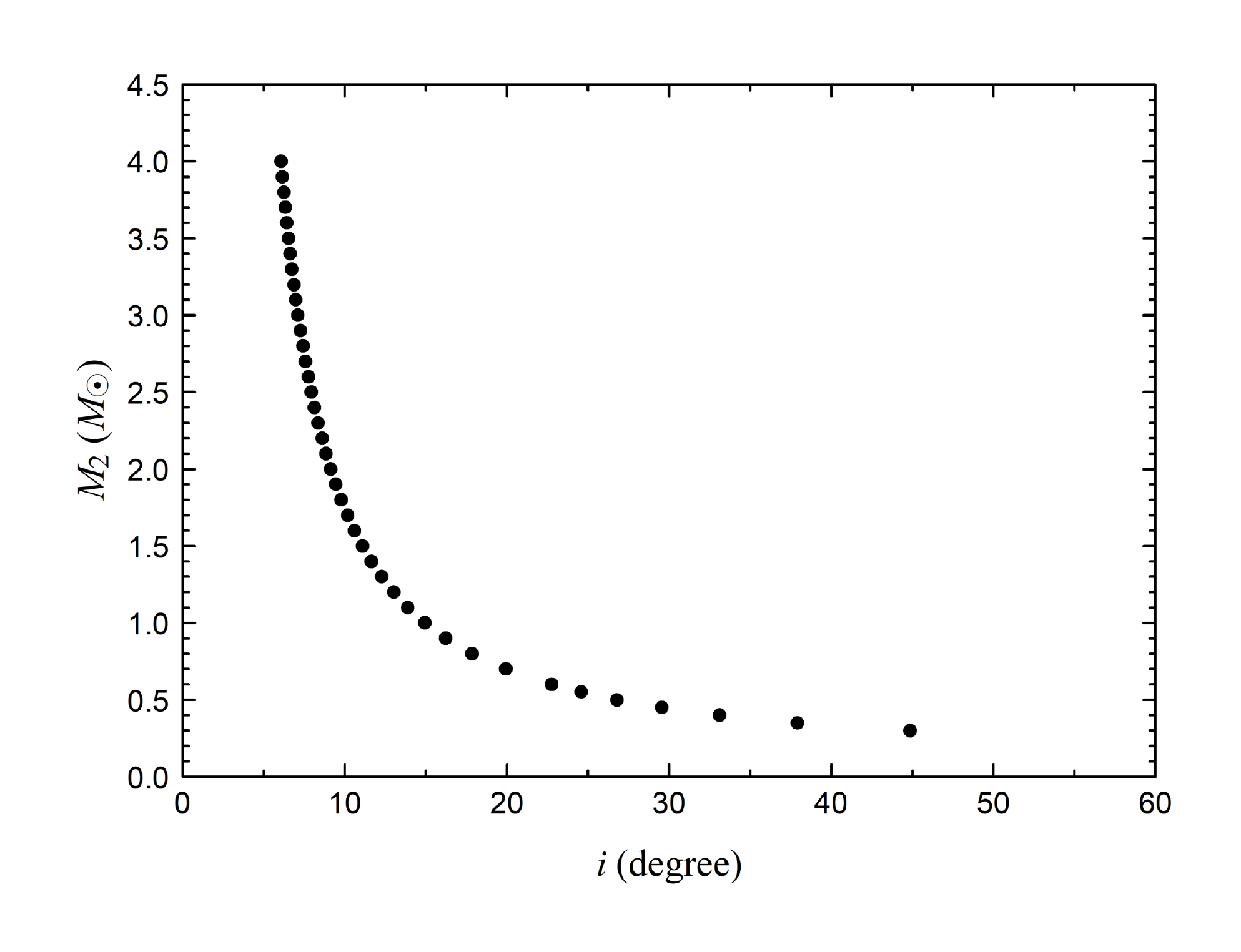}
\caption{Possible conbinations between the orbital inclination (\emph{i}) and the mass of secondary (\emph{M}$_2$) from the mass fuction.}
\label{fig:M2-inclination}
\end{figure}

Based on the fact that the flux we observed is mainly from the primary component, the luminosity (\emph{L} = 10.774 \emph{L}$_\odot$) provided by $\emph{Gaia}$ is approximately that of the primary. By the revised mass-luminosity relation \citep{2015AJ....149..131E}, the mass of the primary component is estimated to be \emph{M}$_1$ = 1.73(3) \emph{M}$_\odot$. From the mass function, we have derived all possible combinations between the orbital inclination (\emph{i}) and the mass of the secondary (\emph{M}$_2$) shown in Figure \ref{fig:M2-inclination}. For \emph{i} $<$ 15$^{\circ}$ (\emph{L}$_2$ $>$ 1 \emph{L}$_\odot$), a double-lined spectroscopic binary would have been found. All other solutions result in a low-mass companion star. Moreover, the orbital inclination \emph{i}$_{orb}$ cannot reach high values, since there are no eclipses seen in the light curves of KIC 10417986 (see subsection \ref{subsection:Fourier analysis}). See section \ref{section:DISCUSSION} for a detailed discussion.

\subsection{Determination of the atmospheric parameters}

\begin{figure}
\centering
\includegraphics[width=0.9\textwidth]{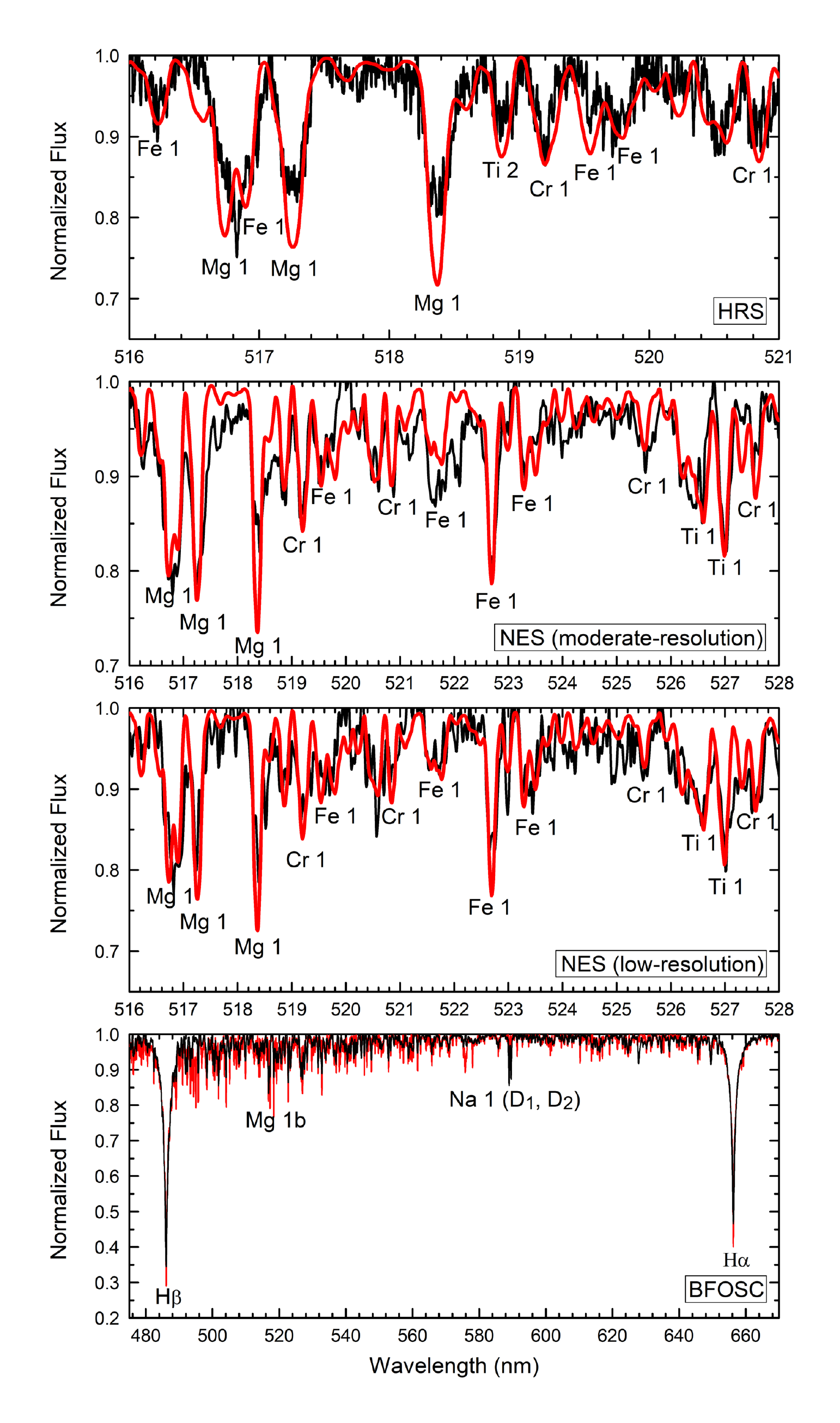}
\caption{Comparisons of the fitted spectra (red line) and the observed spectra of KIC 10417986 (black line). The parts of the spectra observed by HRS, NES (moderate-resolution), NES (low-resolution), and BFOSC are shown from the top to the bottom panels, respectively.}
\label{fig:parameter}
\end{figure}

We derive the atmospheric parameters (i.e. effective temperature \emph{T}$_{eff}$, surface gravity log \emph{g}, metallicity [M/H], and rotation velocity \emph{v}sin\emph{i}) using the synthetic spectra fitting technique provided by the code iSpec \citep{2014A&A...569A.111B, 2019MNRAS.486.2075B}. The synthetic spectra are calculated with the radial transfer code SPECTRUM \citep{1994AJ....107..742G}, the plane-parallel model atmospheric ATLAS9.Castelli (3500 K $\le$ \emph{T}$_{eff}$ $\le$ 8750 K, 0.0 dex $\le$ log \emph{g} $\le$ 5.0 dex, and -5.00 dex $\le$ [M/H] $\le$ 1.00 dex) \citep{2005MSAIS...8...14K}, the solar abundance of \cite{1998SSRv...85..161G}, and the Vienna Atomic Line Database \citep[VALD;][]{1995A&AS..112..525P} line list.

To improve the signal-to-noise ratio, we combine the spectra after correcting the radial velocity on the same observation night (Except for the data on November 25, 2021, because of the very low signal-to-noise ratio), with wavelengths from 480 to 650 {\rm \AA} for HRS, 450 to 650 {\rm \AA} for NES (moderate-resolution), and 450 to 750 {\rm \AA} for NES (low-resolution), respectively. Then we clean the regions of the spectra that may be affected by telluric lines, and degrade the resolution to 80000 and 18000 for HRS and NES, respectively. Finally, we use 3 combined higher resolution spectra and 18 BFOSC lower resolution spectra to derive the atmospheric parameters.

In the iteration process, we set the effective temperature \emph{T}$_{eff}$, surface gravity log \emph{g}, metallicity [M/H], micro-turbulence velocity \emph{v$_{mic}$}, rotation velocity \emph{v}sin\emph{i}, and resolution R as the free parameters, and use the parameters given by the Spitzer Kepler Survey catalog \citep{2021ApJS..254...11W} (see Table \ref{tab:stellar parameters}) as the initial values of \emph{T}$_{eff}$, log \emph{g}, and [M/H]. The macro-turbulence velocity \emph{v$_{mac}$} follows an empirical relation of \cite{2019KPCB...35..129S}, and the limb darkening coefficient is fixed to be 0.6 \citep{1998A&A...333..338H, 2019MNRAS.486.2075B, 2020A&A...636A..85S}. We finally obtain 21 sets of atmospheric parameters by comparing the calculated spectra to the observed spectra. The values of \emph{T}$_{eff}$ = 7411 $\pm$ 187 K and log \emph{g} = 4.2 $\pm$ 0.3 dex are their corresponding mean values derived from the 21 spectra. The values of [M/H] = 0.08 $\pm$ 0.09 dex and \emph{v}sin\emph{i} = 52 $\pm$ 11 km/s are mean values derived from 3 combined higher resolution spectra, as they can not be determined with high precision from the lower resolution spectra. The parts of the spectra fitting to HRS, NES (moderate-resolution), NES (low-resolution), and BFOSC are shown from the top to bottom panel in Figure \ref{fig:parameter}, respectively.

\subsection{Fourier analysis}{\label{subsection:Fourier analysis}}

\begin{figure*}
\centering
\includegraphics[width=1\textwidth]{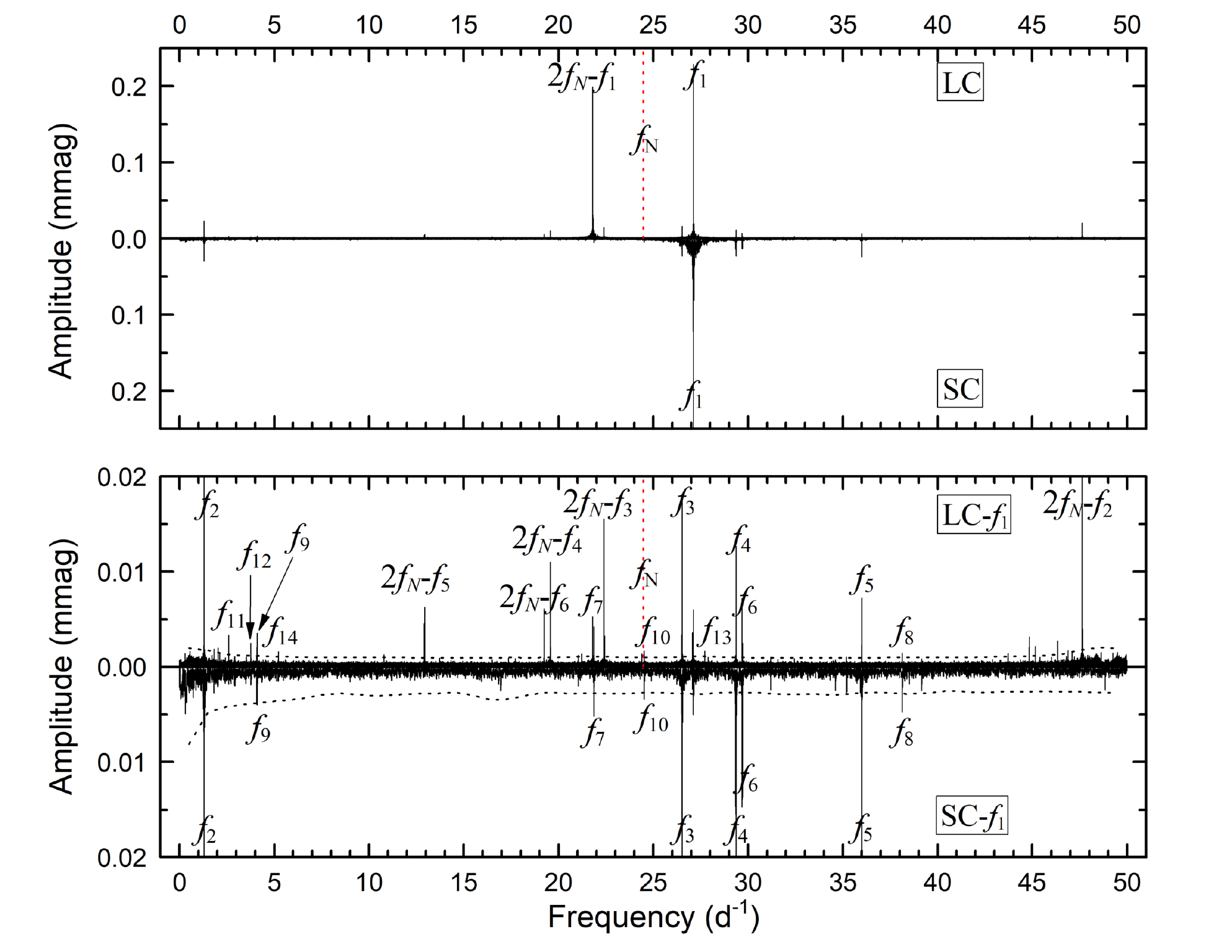}
\caption{Top panel: the Fourier amplitude spectrum of the original LC and SC light curves, in which the dominant frequency is labeled with $f{_1}$. Bottom panel: the Fourier amplitude spectrum of the LC and SC light curves residuals after pre-whitening $f{_1}$, in which the frequencies except $f{_1}$ are indicated. The dotted red lines denote the LC Nyquist frequency labeled with $f{_N}$, and the dotted black lines indicate the 5.6 $\sigma$ detection thresholds of the LC and SC local noise levels, respectively.}
\label{fig:frequency}
\end{figure*}

\begin{table*}
\centering
\caption{List of frequencies detected from LC and SC data of KIC 10417986.}
\label{tab:frequencies}
\setlength{\tabcolsep}{1mm}{
\begin{tabular}{ccccccccc}
\hline
\hline
ID        &Frequency (LC)&Amplitude (LC)&S/N (LC)&Frequency (SC)&Amplitude (SC)&S/N (SC)&$P{_{pul}}$/$P{_{orb}}$&Comment               \\
          &(cd$^{-1}$)   &(mmag)        &        & (cd$^{-1}$)  & (mmag)       &        &                       &                      \\
\hline
$f{_1}$   &27.125672(3)  &0.230(3)      &1180.1  & 27.12568(2)  &  0.407(2)    &  734.8 & 0.044 & \emph{p}-mode        \\
$f{_2}$   &1.303334(8)   &0.0247(5)     &  75.8  & 1.3038(1)    &  0.0303(6)   &   30.6 &  ...  & rotation?            \\
$f{_3}$   &26.532997(9)  &0.0179(4)     &  94.0  & 26.5329(1)   &  0.0294(6)   &   53.3 & 0.045 & \emph{p}-mode        \\
$f{_4}$   &29.36872(1)   &0.0128(4)     &  73.6  & 29.3687(1)   &  0.0243(7)   &   49.2 & 0.040 & \emph{p}-mode        \\
$f{_5}$   &36.00084(2)   &0.0074(4)     &  43.4  & 36.0010(1)   &  0.0240(6)   &   49.0 & ...   & $f{_3}$+8$f{_{orb}}$ \\
$f{_6}$   &29.69563(2)   &0.0073(3)     &  41.7  & 29.6954(3)   &  0.0148(6)   &   30.4 & 0.040 & \emph{p}-mode        \\
$f{_7}$   &21.87028(3)   &0.0043(3)     &  22.0  & 21.8705(7)   &  0.0054(7)   &   10.0 & 0.054 & \emph{p}-mode        \\
$f{_8}$   &38.14077(8)   &0.0015(4)     &   8.5  & 38.1399(8)   &  0.0048(6)   &    9.2 & 0.031 & \emph{p}-mode        \\
$f{_9}$   &4.09286(3)    &0.0036(3)     &  18.6  & 4.0922(9)    &  0.0041(6)   &    6.3 & ...   & 4$f{_8}$$-$5$f{_6}$?   \\
$f{_{10}}$&24.527(3)     &0.0017(2)     &   9.8  & 24.53(9)     &  0.0034(4)   &    6.9 & ...   & $f{_1}$$-$2$f{_2}$     \\
\hline
$f{_{11}}$& 2.60698(4)   &0.0033(3)     & 15.1   & ...          &  ...         & ...    & ...   & 2$f{_2}$             \\
$f{_{12}}$& 3.775(9)     &0.0025(3)     & 12.9   & ...          &  ...         & ...    & ...   & 2$f{_2}$+$f{_{orb}}$ \\
$f{_{13}}$& 27.711(5)    &0.0017(2)     &  8.6   & ...          &  ...         & ...    & ...   & $f{_3}$+$f{_{orb}}$  \\
$f{_{14}}$& 5.21475(9)   &0.0016(3)     &  8.4   & ...          &  ...         & ...    & ...   & 4$f{_2}$             \\
\hline
\end{tabular}
}
\end{table*}

The Fourier analysis is performed with the software Period04 v.1.2.0 \citep{2005CoAst.146...53L} to LC and SC data of KIC 10417986, respectively. A multi-frequency non-linear least-squares fit is used to extract significant frequencies of each light curve with the equation:
\begin{equation}
\Delta m = \sum_{i} A_{i}sin[2\pi \left ( f_{i} t + \phi _{i}\right )]
\label{eq:least-squares fit}
\end{equation}
where \emph{A}$_i$ is the amplitude, \emph{f}$_i$ is the frequency, and $\phi_i$ is the phase. We search for significant peaks up to the LC and SC Nyquist frequencies (24.5 cd$^{-1}$ and 734 cd$^{-1}$, respectively) in the first run, but no peaks are found beyond the frequency of 50 cd$^{-1}$ in SC data. So, we limit the lower and upper bound of the frequency to 0 and 50 cd$^{-1}$ for both LC and SC data, and a standard pre-whitening procedure \citep{2019PASP..131f4202Z} is performed to subtract frequencies from light curves until a 5.6 $\sigma$ limit (5.6 $\sigma$ meaning the risk of a false detection less than 1/10000, recommended by \citealt{2016A&A...585A..22Z}). The Nyquist alias frequencies in LC data are identified by cross-matching with the frequencies in SC data \citep{2013MNRAS.430.2986M}.

A total of 14 and 10 significant frequencies are detected from LC and SC light curves of KIC 10417986, respectively. The frequency values of LC and SC data with respective amplitude and signal-to-noise ratio (S/N) are listed in Table \ref{tab:frequencies}. The corresponding uncertainties of frequencies and amplitudes are derived by using the Monte Carlo Simulation in Period04, and the S/N is calculated in a box size of 2 cd$^{-1}$ centered on the extracted frequency. All of the frequencies and LC Nyquist alias frequencies are labeled in Figure \ref{fig:frequency}. The dotted red lines denote the LC Nyquist frequency labeled with $f{_N}$, and the dotted black lines indicate the 5.6 $\sigma$ detection thresholds.

Among these significant frequencies, we identify combinations $f{_i}$ using the formula: $\mid$ $f{_i}$ - (\emph{m} $\times$ $f{_j}$ + \emph{n} $\times$ $f{_k}$) $\mid$ $<$ $\varepsilon$, where \emph{m} and \emph{n} are integers between -5 and 5, $f{_j}$ and $f{_k}$ are independent frequencies, and $\varepsilon$ is the Rayleigh resolution ($\varepsilon$ = {1.5}/{$\Delta$T} \citep{1978Ap&SS..56..285L}, $\varepsilon$ $\approx$ 0.0013 and 0.017 cd$^{-1}$ for LC and SC data, respectively) \citep{2012AN....333.1053P, 2015MNRAS.450.3015K}. Since there is an orbital period threshold of $\sim$13 days below which the pulsation properties are affected by the binarity \citep{2015ASPC..496..195L}, the possible combinations with the orbit harmonics ($f{_{orb}}$ = 1.18350(3) cd$^{-1}$) are also identified. A combination is accepted if the difference is less than Rayleigh resolution. The results are listed in the last column of Table \ref{tab:frequencies}.

The harmonics of $f{_2}$ ($f{_{11}}$ = 2$f{_2}$ and $f{_{14}}$ = 4$f{_2}$) are detected in LC data, suggesting that $f{_2}$ may be the orbital or rotational frequency \citep{2013MNRAS.431.2240B, 2014ApJS..211...24M}. Assuming that $f{_2}$ is an orbital frequency, the corresponding orbital period \emph{P} = 2/$f{_2}$ $\approx$ 1.5342 d. However, this period can not match the observed radial velocities. Therefore, $f{_2}$ is probably due to rotation, which will be discussed in section \ref{section:DISCUSSION}. We identify $f{_9}$ as a combination frequency with a difference slightly above the Rayleigh resolution of LC data. However, a higher combination order (order $>$ 2) will increase the probability of a chance of coincidence with an independent frequency \citep{2012AN....333.1053P, 2016MNRAS.459.1097B}. Therefore, the possibility that $f{_9}$ is a \emph{g}-mode pulsation frequency cannot be ruled out. It is worth noting that $f{_1}$ shows a non-sinusoidal light variability with an amplitude one order of magnitude higher than that of other frequencies. We find that $f{_1}$ is approximately 23 times the orbital frequency, indicating that the pulsation characteristics of $f{_1}$ are possibly affected by the orbital effect.

The results from the spectral and Fourier analysis show that the primary component of KIC 10417986 match well with the profiles of $\delta$ Scuti stars (e.g. effective temperature, luminosity, and period range) \citep{2000ASPC..210....3B}. \cite{2013ApJ...777...77Z} first deduced the theoretical relation between the pulsation and orbital period for eclipsing binaries containing $\delta$ Sct components, and suggest that the pulsation could be of \emph{p}-mode if the \emph{P}$_{pul}$/\emph{P}$_{orb}$ ratio is below 0.07. In this work, we calculate the \emph{P}$_{pul}$/\emph{P}$_{orb}$ ratios of six independent frequencies ($f{_1}$, $f{_3}$, $f{_4}$, $f{_6}$, $f{_7}$, and $f{_8}$) with an uncertainty of $10{^{-6}}$. All of the ratios are below 0.07, this indicate that these frequencies probably belong to \emph{p}-mode pulsations. The results are listed in the last two columns of Table \ref{tab:frequencies}. 

\section{DISCUSSION}{\label{section:DISCUSSION}}

\begin{figure}
\centering
\includegraphics[width=0.8\textwidth]{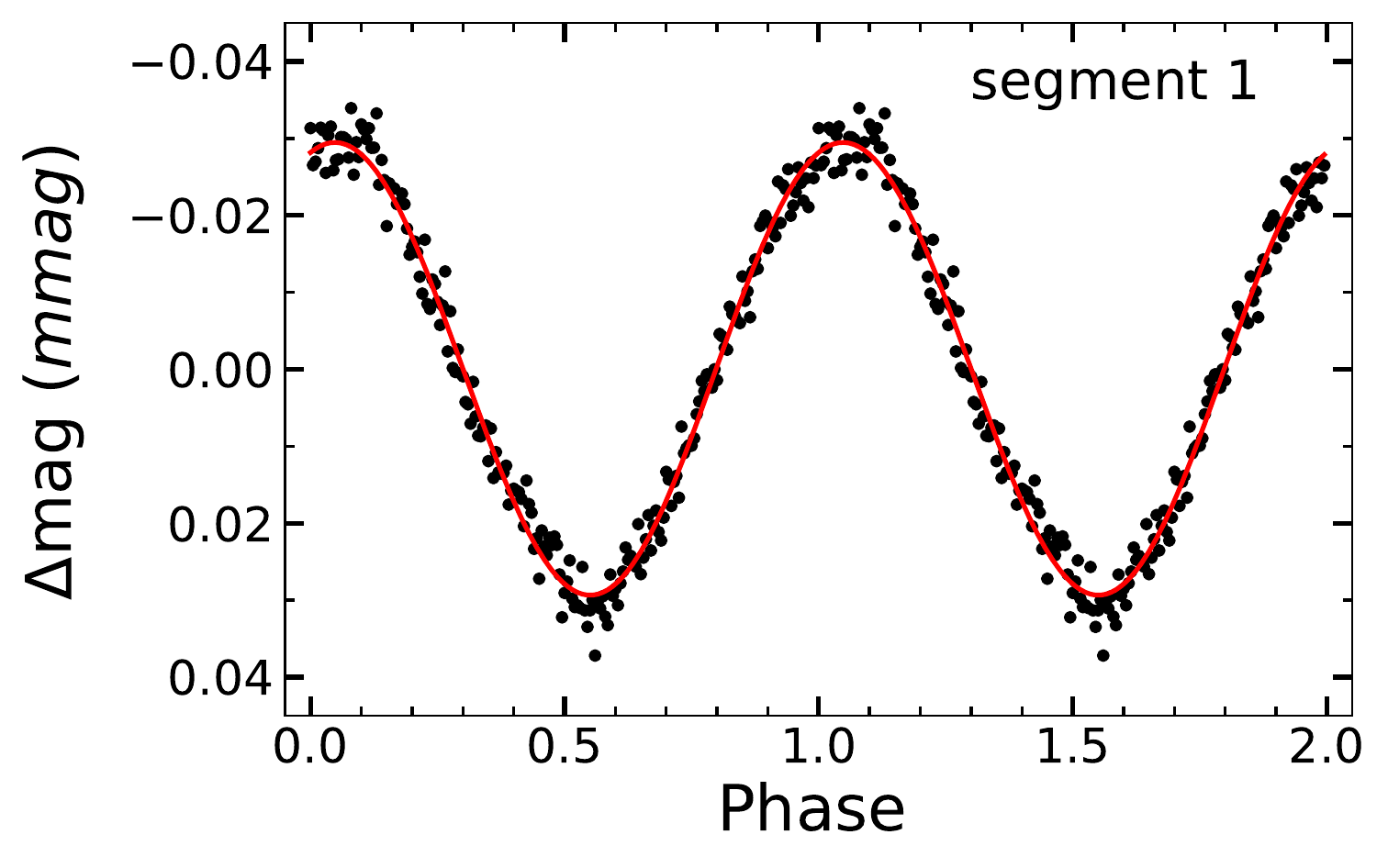}
\caption{The observed light curve (black-filled circles) and the synthetic light curve from the Fourier models (red line) for segment 1.}
\label{fig:ellipsoidal}
\end{figure}

\begin{figure*}
\centering
\includegraphics[width=1\textwidth]{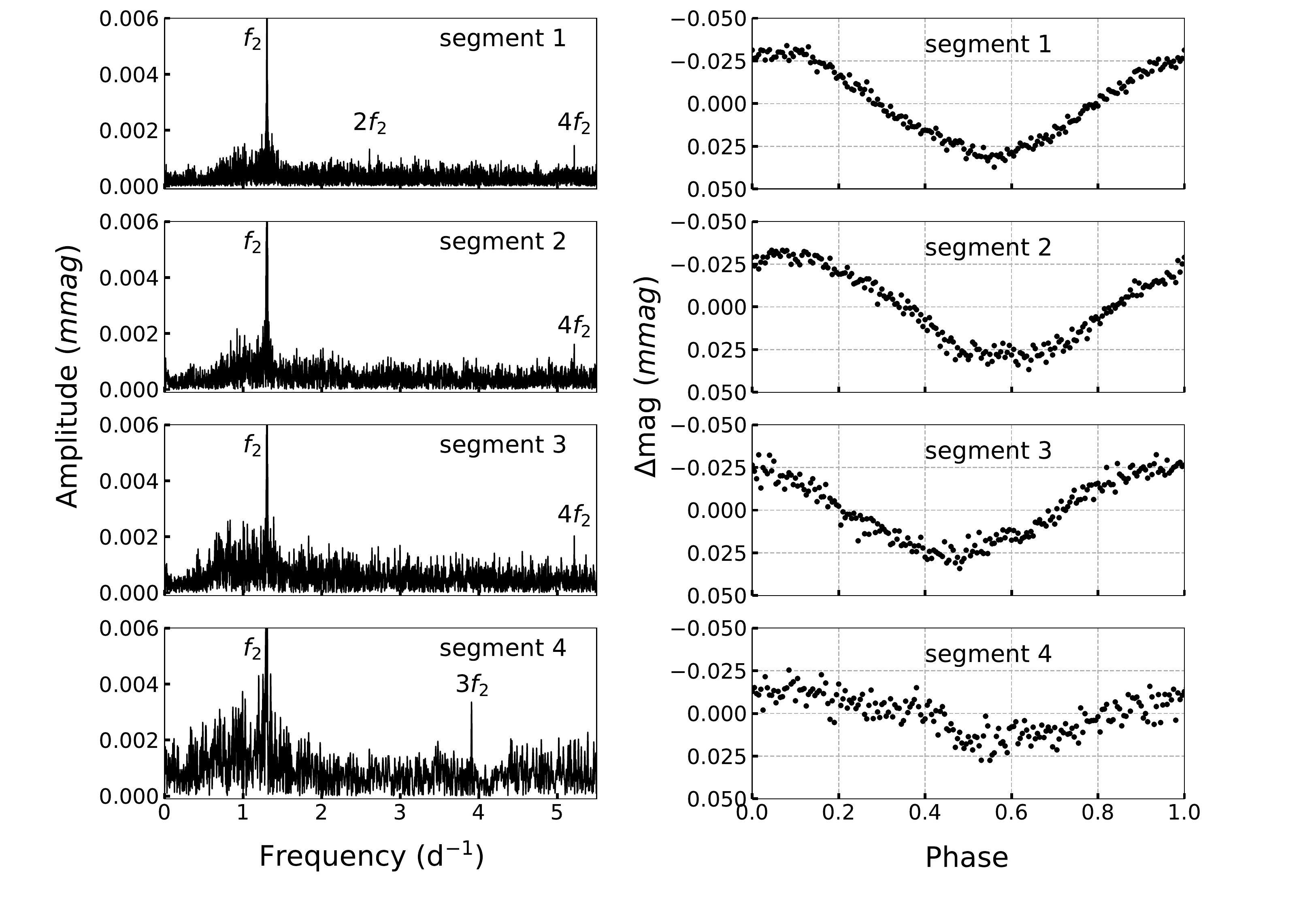}
\caption{A comparison of the amplitude and phase of $f{_2}$ in the four segments.}
\label{fig:rotation}
\end{figure*}

According to the radial velocities and light curves analysis of KIC 10417986, the frequency $f{_2}$, which is non-sinusoidal, is probably due to rotation rather than the case of \emph{g}-mode pulsation, since harmonics are generally expected at high amplitudes \citep{2013MNRAS.431.2240B, 2015MNRAS.450.3015K}, yet the amplitude of $f{_2}$ is only $\sim$0.03 mmag. We first check if the light variation of $f{_2}$ is caused by the ellipsoidal effect, which occurs in non-eclipsing binary stars and especially close binaries \citep{1985ApJ...295..143M, 1985Ap&SS.117...69B, 1993ApJ...419..344M}, by modeling the light curves with the Fourier analysis based on the methods of \cite{2011MNRAS.415.3921F} and \cite{2012ApJ...746..185F}. The formula of Fourier analysis was applied to fit the light curves of segment 1 (see Figure \ref{fig:LC-SC-ligthcurves}) as follows:
\begin{equation}
\label{eq:ellipsoidal}
\resizebox{,6\hsize}{!}{$L(\theta) = A_{0} + A_{1}cos(\theta) + A_{2}cos(2\theta) + B_{1}sin(\theta) + B_{2}sin(2\theta)$}.
\end{equation}
The synthetic light curve from the Fourier models is shown in Figure \ref{fig:ellipsoidal}, and the derived Fourier coefficients are as follows: $A_{0}$ = $-$0.00013980(2), $A_{1}$ = $-$0.02802546(3), $A_{2}$ = $-$0.00000041(3), $B_{1}$ = $-$0.00891906(3), and $B_{2}$ = 0.00013332(3). The results show that the $cos(\theta)$ term is the most dominant. Thus, it is obvious that the light variation of $f{_2}$ is mainly due to the magnetic activity or other effects (e.g. reflection effect) rather than the ellipsoidal effect.

To understand the nature of $f{_2}$, we divide LC data of KIC 10417986 into four segments (see Figure \ref{fig:LC-SC-ligthcurves}), which have a good continuous sampling. Firstly, all frequencies except frequency $f{_2}$ are pre-whitened from the data in each segment. Then, the residual light curves are used for the Fourier analysis. The Fourier amplitude spectrum of the residuals of each segment is shown in the left panels of Figure \ref{fig:rotation}. The second and fourth harmonics of frequency $f{_2}$ are detected in the residuals of segment 1, only the fourth harmonic in segment 2 and segment 3, and the third harmonic in segment 4. The phase-folded light curves of $f{_2}$ for each segment are also shown in the right panels of Figure \ref{fig:rotation}. Note that the data has been averaged using binning, and the number of bins is 200. It can be seen that $f{_2}$ exhibits an amplitude and phase variation over a time span of about four years, indicating that the variations of $f{_2}$ are probably caused by starspots. In this case, the amplitudes of the harmonic mainly depend on the inclination of the rotation axis and the position of the spots and will sometimes be low or absent \citep{2013MNRAS.431.2240B}. \cite{2017AJ....154..250L} find that some of short-period binaries in 816 EBs with starspots modulations have rotation periods 13\% slower than the orbital period, which is probably due to the surface differential rotation of starspots. However, we find the orbital period (0.84495 days) is slightly slower than the starspots rotation period (1/$f{_2}$ = 0.76726 days). It could be the same reason but the starspots locate in the fast rotation regions \citep{2022RAA....22a5016N}. The surface differential rotation of stars is also found in many other works \citep{1996ApJ...466..384D, 2000MNRAS.314..162B, 2005AN....326..287W, 2012IAUS..286..268K, 2018MNRAS.474.5534O}.

According to the results of \cite{2017AJ....154..250L}, binary stars with an orbital period less than 2 days are almost synchronous. Moreover, the synchronization time-scale must be shorter than the circularization time-scale due to the orbit's angular momentum being generally larger than that of the stars \citep{1975A&A....41..329Z, 2008EAS....29...67Z}. As KIC 10417986 is a circular orbit binary system, we propose here that the rotation period should be the orbital period 0.84495 days of this star, and assume that the \emph{i}$_{rot}$ $\approx$ \emph{i}$_{orb}$. Then, the derived equatorial rotation velocity \emph{v} is about 120 km/s from the estimated radius of 2.0(1) \emph{R}$_\odot$. The \emph{v}sin\emph{i} of 52 $\pm$ 11 km/s has been obtained from the synthetic spectra fitting. Thus, the orbital inclination \emph{i}$_{orb}$ is estimated to be 26 $\pm$ 6$^{\circ}$, and subsequently, the mass of the secondary \emph{M}$_2$ is estimated to be $0.52^{+0.18}_{-0.09}$ \emph{M}$_\odot$, which is a late-K to early-M type star.

\begin{figure}
\centering
\includegraphics[width=1\textwidth]{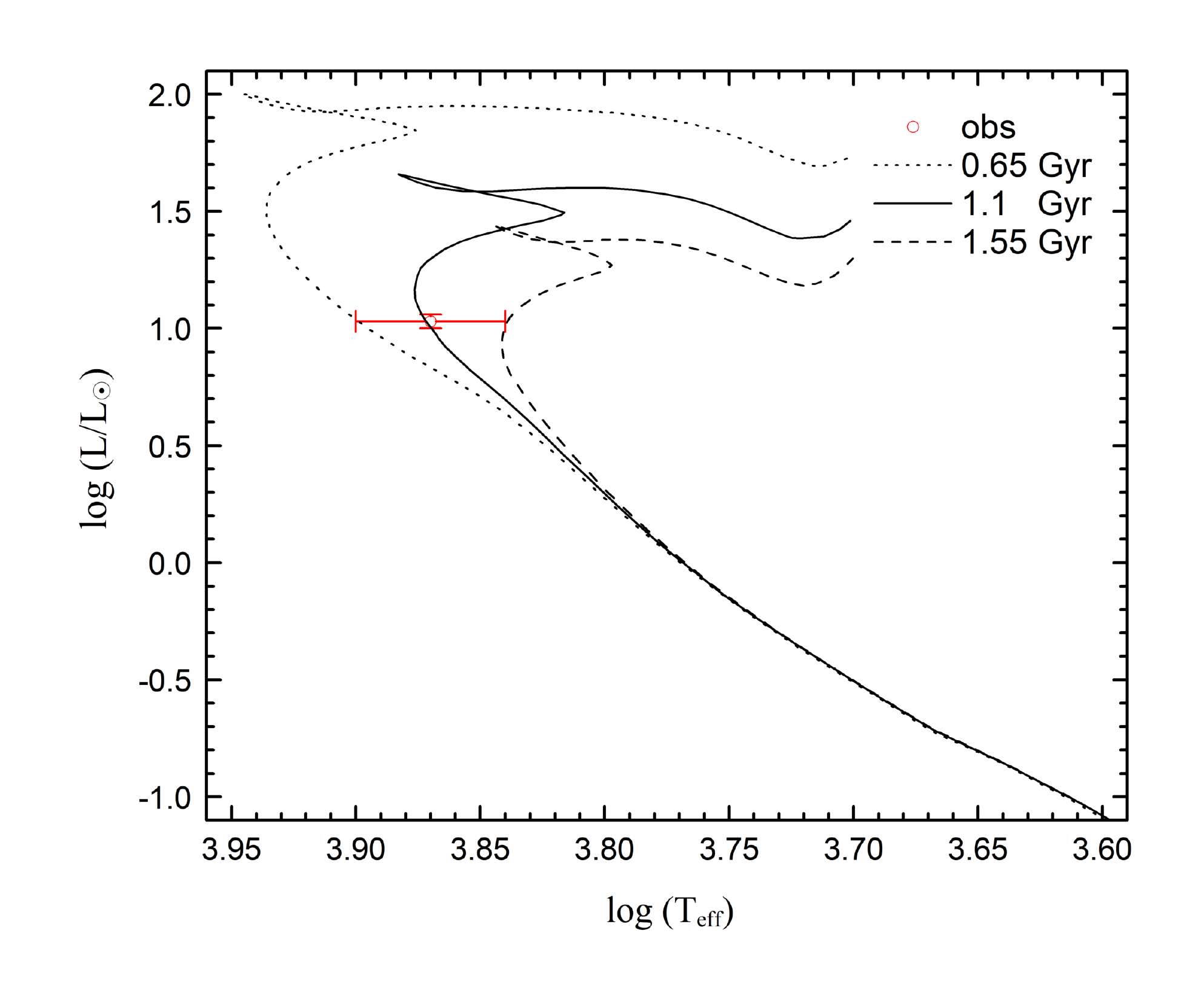}
\caption{PARSEC isochrones for KIC 10417986 with [M/H] = 0.08 dex. The location of the primary component is labeled with a red open circle with $\pm$3 $\sigma$ errorbars. The dotted, solid, and dashed lines are isochrones for age of 0.65 Gyr, 1.1 Gyr, and 1.55 Gyr, respectively.}
\label{fig:age}
\end{figure}

KIC 10417986 is a short-period circular orbit binary system with a semi-major axis \emph{a} of about 5 \emph{R}$_\odot$. We assume that the two components have not undergone material exchange. Then, the age of the system is estimated by comparing the observed temperature and luminosity of the primary with a web interface the PAdova and TRieste Stellar Evolution Code\footnote{http://stev.oapd.inaf.it/cgi-bin/cmd\_3.4} \citep[PARSEC, version 1.2S;][]{2012MNRAS.427..127B} isochrones. We use the approximation [M/H] = log(\emph{Z}/\emph{X}) $-$ log(\emph{Z}/\emph{X})$_\odot$ for PARSEC tracks, where the [M/H] = 0.08 dex has been derived from the synthetic spectra fitting, and the present solar metal content are \emph{Z}$_\odot$ = 0.0152, and \emph{Y} = 0.2485 + 1.78\emph{Z}. Figure \ref{fig:age} shows the location of the primary component with a $\pm$3 $\sigma$ credible region in the H-R diagram. Three isochrones of age = 0.65, 1.1, and 1.55 Gyr are plotted with the dotted, solid, and dashed lines, respectively. The age of KIC 10417986 is estimated to be 1.1 $\pm$ 0.45 Gyr, and the primary has evolved to the late stages of the main sequence.

\section{SUMMARY}{\label{section:SUMMARY}}
In this paper, the ground-based spectroscopic data and \kep high-precision photometric data are combined to study the non-eclipsing binary KIC 10417986. A total of 29 single-lined spectra were observed with the 2.16-m telescope at the Xinglong Station and the 1.2-m telescope at Nanshan Station. The orbital parameters of $P{_{orb}}$ = 0.84495 $\pm$ 0.00002 d, \emph{K}$_1$ = 29.7 $\pm$ 1.5 kms$^{-1}$, and $\gamma$ = $-$18.7 $\pm$ 1.7 kms$^{-1}$ are obtained by using the rvfit code, and the derived quantities are $a_1\sin i$ = 0.50 $\pm$ 0.02 $R_\odot$, $f(m_1,m_2)$ = 0.0023 $\pm$ 0.0003 $M_\odot$. The atmospheric parameters of the primary are determined by the synthetic spectra fitting technique, where the \emph{T}$_{eff}$ = 7411 $\pm$ 187 K and log \emph{g} = 4.2 $\pm$ 0.3 dex are derived from the 3 combined higher resolution and 18 BFOSC lower resolution spectra, while the [M/H] = 0.08 $\pm$ 0.09 dex and \emph{v}sin\emph{i} = 52 $\pm$ 11 km/s from only the 3 combined higher resolution spectra. There are 14 frequencies extracted from the Fourier analysis of LC and SC data with S/N $<$ 5.6, of which $f{_2}$ is probably the rotational frequency due to the starspots, six independent pulsation frequencies in the high-frequency region ($f{_1}$, $f{_3}$, $f{_4}$, $f{_6}$, $f{_7}$, and $f{_8}$) probably belong to the $p$-mode with the \emph{P}$_{pul}$/\emph{P}$_{orb}$ ratios below 0.07 and the rest are possibly the combinations or harmonic frequencies. From the estimated mass and radius of the primary, and the observed \emph{v}sin\emph{i}, the derived orbital inclination is 26 $\pm$ 6$^{\circ}$, and the mass of the secondary is $0.52^{+0.18}_{-0.09}$ \emph{M}$_\odot$, which should be a late-K to early-M type star. Finally, we estimate the age of the binary star to be about 1.1 $\pm$ 0.45 Gyr.

\begin{acknowledgements}
Thanks to the referee for helpful comments and the editor for carefully revision the manuscript. We acknowledge the support from the National Natural Science Foundation of China (NSFC) through grants 11403088, 11873081, U2031209, 11833002, 12090040, 12090042 and 12003020. This research is supported by the Nanshan 1.2-m telescope of Xinjiang Astronomical Observatory, Chinese Academy of Sciences. We also acknowledge the support of the staff of the Xinglong 2.16-m telescopes. This work was partially supported by the Open Project Program of the Key Laboratory of Optical Astronomy, National Astronomical Observatories, Chinese Academy of Sciences. This paper includes data collected by the \kep mission. Funding for the \kep mission is provided by the NASA Science Mission directorate. All of the \kep data presented in this paper were obtained from the Mikulski Archive for Space Telescopes (MAST). This research has made use of data from the European Space Agency (ESA) mission Gaia (https://www.cosmos.esa.int/gaia), processed by the Gaia Data Processing and Analysis Consortium (DPAC, https://www.cosmos.esa.int/web/gaia/dpac/consortium). Funding for the DPAC has been provided by national institutions, in particular the institutions participating in the Gaia Multilateral Agreement. 
\end{acknowledgements}

\label{lastpage}

\end{document}